\documentclass[10pt, conference]{IEEEtran}
\AtBeginDocument{%
  \providecommand\BibTeX{{%
    \normalfont B\kern-0.5em{\scshape i\kern-0.25em b}\kern-0.8em\TeX}}}

\usepackage{hyperref}
\usepackage{lipsum} % For dummy text
\usepackage{amsmath,amssymb,amsfonts}
\usepackage{graphicx}
\graphicspath{{Figure/}}
\usepackage{textcomp}
\usepackage{xcolor}
\usepackage[caption=false,font=normalsize,labelfont=sf,textfont=sf]{subfig}
\usepackage{url}
\usepackage{amsmath}
\usepackage{booktabs}
\usepackage{tabularx}
\usepackage{algorithm}
\usepackage[noend]{algpseudocode}
\usepackage{titlesec}
\usepackage{enumitem}   
\usepackage{booktabs} % For better table formatting
\usepackage{multirow}
\usepackage{array}     % For m{} column formatting
\usepackage{amsmath}
\usepackage{amsfonts}
\usepackage{multicol}  % for two column algorithm
\usepackage{tcolorbox}
\usepackage{xcolor}
\usepackage[numbers,sort&compress]{natbib}

\usepackage{listings}
\usepackage{adjustbox}
\usepackage{xfrac}
\usepackage{pgfplots}
\usepackage{pgfplotstable}
\pgfplotsset{compat=1.18}
\usepackage{tikz}
\renewcommand{\arraystretch}{0.8} % Reduce row spacing globally

\begin{document}
\pagestyle{plain}

%\fancyhead{} % This line clears the header entirely
%\pagestyle{empty} % removes running headers

%\usepackage{fancyhdr}
% New definitions
% \algnewcommand\algorithmicswitch{\textbf{switch}}
% \algnewcommand\algorithmiccase{\textbf{case}}
% \algnewcommand\algorithmicassert{\texttt{assert}}
% \algnewcommand\Assert[1]{\State \algorithmicassert(#1)}%
% % New "environments"
% \algdef{SE}[SWITCH]{Switch}{EndSwitch}[1]{\algorithmicswitch\ #1\ \algorithmicdo}{\algorithmicend\ \algorithmicswitch}%
% \algdef{SE}[CASE]{Case}{EndCase}[1]{\algorithmiccase\ #1}{\algorithmicend\ \algorithmiccase}%
% \algtext*{EndSwitch}%
% \algtext*{EndCase}%

%\pagestyle{empty}
%\title{\fontsize{18}{22}\selectfont Your Title Goes Here}
\title{\fontsize{18}{22}\selectfont Silentflow: Leveraging Trusted Execution for Resource-Limited MPC via Hardware-Algorithm Co-design\\ %\vspace{-0.5in}
}

\author{%
\IEEEauthorblockN{
Zhuoran Li\IEEEauthorrefmark{1},
Hanieh Totonchi Asl\IEEEauthorrefmark{1},
Ebrahim Nouri\IEEEauthorrefmark{1}\\
Yifei Cai\IEEEauthorrefmark{1},
Danella Zhao\IEEEauthorrefmark{1}
}
\IEEEauthorblockA{\IEEEauthorrefmark{1}\textit{Electrical and Computer Engineering}, \textit{University of Arizona}, Tucson, United States\\
\{zli1122, haniehta, ebinouri, danellazhao\}@arizona.edu}
}

\maketitle

%\IEEEtitleabstractindextext{%
\begin{abstract}

Secure Multi-Party Computation (MPC) offers a practical foundation for privacy-preserving machine learning at the edge, with MPC commonly employed to support nonlinear operations. These MPC protocols fundamentally rely on Oblivious Transfer (OT), particularly Correlated OT (COT), to generate correlated randomness essential for secure computation. Although COT generation is efficient in conventional two-party settings with resource-rich participants, it becomes a critical bottleneck in real-world inference on resource-constrained devices (e.g., IoT sensors and wearables), due to both communication latency and limited computational capacity. To enable real-time secure inference, we introduce Silentflow, a highly efficient Trusted Execution Environment (TEE)-assisted protocol that eliminates communication in COT generation. We tackle the core performance bottleneck—low computational intensity—through structured algorithmic decomposition: kernel fusion for parallelism, Blocked On-chip eXpansion (BOX) to improve memory access patterns, and vectorized batch operations to maximize memory bandwidth utilization. Through design space exploration, we balance end-to-end latency and resource demands, achieving up to 39.51$\times$ speedup over state-of-the-art protocols. By offloading COT computations to a Zynq-7000 SoC, SilentFlow accelerates PPMLaaS inference on the ImageNet dataset under resource constraints, achieving a 4.62$\times$ and 3.95$\times$ speedup over Cryptflow2 and Cheetah, respectively.

\end{abstract}

\begin{IEEEkeywords}
Security \& Privacy, Multiparty Computation, Trusted Execution Environment, FPGA acceleration
\end{IEEEkeywords}

\section{Introduction}\label{Sec:Intro_yifei}

To address privacy concerns in Machine Learning as a Service (MLaaS), Privacy-Preserving MLaaS (PPMLaaS) incorporates cryptographic primitives to safeguard sensitive data~\cite{juvekar2018gazelle, riazi2019xonn, rouhani2018deepsecure, mishra2020delphi, zhang2021gala}. Among these primitives, secure Multi-Party Computation (MPC) has been widely adopted for its effectiveness in handling nonlinear operations~\cite{rathee2020cryptflow2, huang2022cheetah, pang2024bolt, lu2025bumblebee, zhang2024individual}. A core component of MPC protocols is Oblivious Transfer (OT), which plays a critical role in ensuring data privacy between parties.

\begin{figure}[t]
  \centering
  \includegraphics[width =\linewidth]{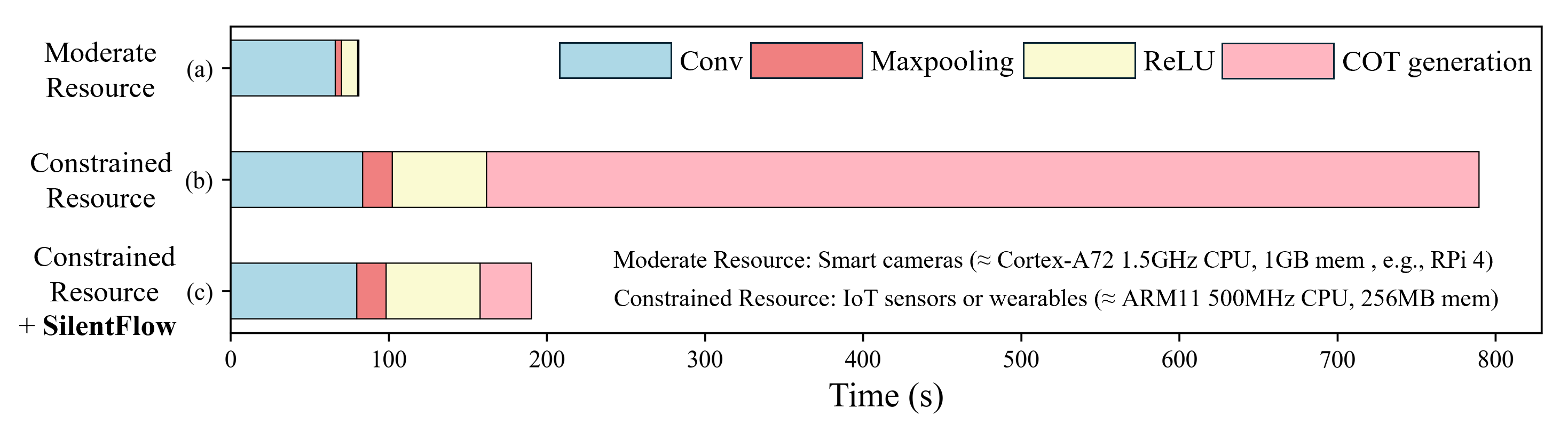}
  \captionsetup{skip=-20pt}
  \caption{Resnet-50 inference performance under 2-Party Cheetah~\cite{huang2022cheetah} framework with a network condition of 200Mbps bandwidth, 50ms latency}
  \label{fig:prelminary_memory_network} 
\end{figure}

Over time, OT protocols have evolved significantly to improve efficiency and scalability. Recent advancements, such as Silent OT~\cite{boyle2019efficient1} and the Ferret protocol~\cite{yang2020ferret}, demonstrate superior performance compared to classic OT protocols like IKNP~\cite{kolesnikov2013improved}, particularly in throughput and communication overhead. While the cost of generating usable OTs is often negligible~\cite{boyle2018compressing,boyle2019efficient,boyle2019efficient1,yang2020ferret,boyle2022correlated,boyle2023oblivious,couteau2021silver} in two-party computation (2PC) settings—where both client and server run on resource-rich platforms such as workstations—it becomes a major bottleneck in real-world inference scenarios involving resource-constrained client devices, as shown in Fig.~\ref{fig:prelminary_memory_network}.

\textbf{Motivations and Challenges.} In real-world PPMLaaS deployments—particularly at the edge—clients often run on resource-constrained hardware such as IoT sensors, wearables, or portable medical devices that have limited memory and computational power~\cite{zheng2019challenges, aminifar2024privacy}, which introduces two critical challenges. First, limited memory necessitates splitting inference into small batches, as there is insufficient storage to retain large intermediate values—particularly correlated oblivious transfers (COTs), which are essential for secure computation of nonlinear operations like ReLU. As a result, even a single nonlinear layer must be executed batch by batch, with fresh COTs generated for each batch. This repeated COT generation intensifies interactive rounds and results in substantial communication overhead. Second, %in current state-of-the-art works~\cite{boyle2019efficient1, yang2020ferret, couteau2021silver}, 
% the state-of-the-art COT generation~\cite{boyle2019efficient1, yang2020ferret, couteau2021silver} is dominated by low arithmetic intensity and high memory traffic, leading to severe performance degradation on embedded devices
since COT generation is dominated by low arithmetic intensity and high global memory traffic, state-of-the-art methods~\cite{boyle2019efficient1,yang2020ferret,couteau2021silver} suffer performance degradation in resource-limited environments—%primarily due to frequent cache misses when working sets exceed the limited on-chip memory, rather than due to raw computation cost. 
not due to computational limitations, but primarily due to frequent cache misses when working sets exceed the limited on-chip memory.
% This causes frequent cache %and memory 
% misses during the generation process, leading to 
These frequent cache misses during the generation process lead to inefficient memory access and increased client-side latency%that delays the overall inference pipeline
. This inefficiency not only increases end-to-end inference %latency 
time but also %leads to a waste of
wastes server-side resources, as servers often % reserve computational capacity and await responses from slow clients.
idle while awaiting responses from slow clients.

%As shown in Fig.~\ref{fig:prelminary_memory_network}(a) and (b), a resnet-50 secure inference is implemented under three different client resources with the current COT protocols (Ferret). Due to the limited 

To address these dual bottlenecks, we propose a set of targeted design optimizations as \textbf{contributions}:

\begin{itemize}
    \item First, we introduce a novel TEE-assisted protocol that eliminates the communication overhead inherent in OT extension. SilentFlow leverages synchronized seeds within each party’s TEE to generate a minimal amount of correlated randomness, effectively bridging the gap between prior interactive approaches~\cite{boyle2019efficient,boyle2019efficient1,yang2020ferret,couteau2021silver}. This allows both parties to locally derive identical pseudorandom values and establish correlated randomness—traditionally requiring interaction—without any communication, achieving up to a 39.51$\times$ speedup over prior approaches~\S\ref{sec:cot_generation}. In contrast to previous methods that delegate extensive computation to the TEE—potentially exposing client inputs—SilentFlow confines the TEE’s role to input-independent COT generation~\cite{bahmani2017secure, choi2019hybrid, wu2022hybrid, zhou2022ppmlac,zhou2022efficient}. As a result, even if the TEE is compromised, client data and model parameters remain secure throughout the end-to-end PPMLaaS inference. 

    \item Second, to address the memory-bound bottleneck in COT generation, we propose a hardware–algorithm co-design that employs structured algorithmic decomposition to optimize memory access patterns and reduce intermediate data movement. By fusing parallel kernels, introducing Blocked On-chip eXpansion (BOX), applying vectorized batch operations, and exploring the design space to balance performance and resource utilization, SilentFlow achieves scalable, low-latency execution with 10.21$\times$ speedup, while using only 160KB of local memory.
    
    \item Finally, we conduct extensive experiments by offloading COT generation to a Zynq-7000 SoC, accelerating secure ResNet-50 inference with ImageNet dataset by 4.62$\times$ and 3.95$\times$ over Cryptflow2~\cite{rathee2020cryptflow2} and Cheetah~\cite{huang2022cheetah}, respectively.

\end{itemize}

\section{Preliminaries}\label{Sec:Preliminaries}

\subsection{Oblivious Transfer}

OT enables a sender to transmit multiple messages such that the receiver learns only the one matching their selection bit, while the sender learns nothing about the receiver’s choice. A complete OT protocol consists of an \textit{input-independent} correlated randomness generation phase and an \textit{input-dependent} oblivious data transfer phase, during which the client’s secret message exchange occurs. The former dominates communication cost, while the latter incurs only linear, information-theoretic cost with respect to the number of messages~\cite{boyle2018compressing}.

\begin{figure}[h!]
  \centering
  \includegraphics[width =0.9\linewidth]{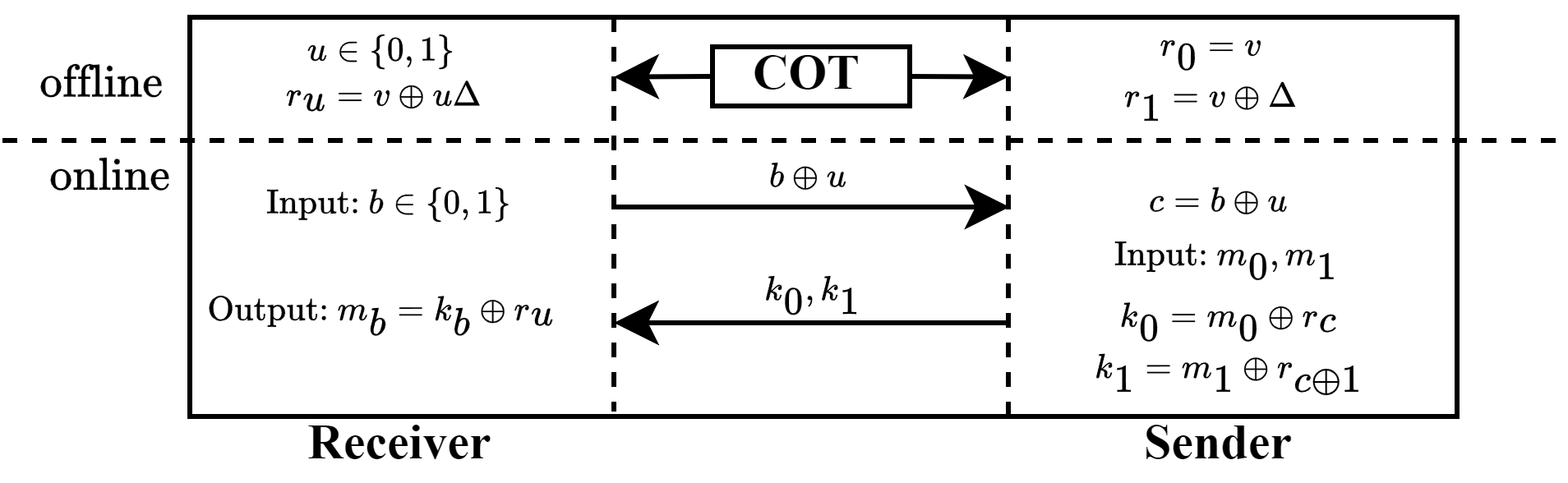}
  %\captionsetup{skip=-15pt}
  \vspace{-0.05in}
  \caption{Oblivious Transfer process}
  \label{fig:cot} 
\end{figure}

As illustrated in Fig.~\ref{fig:cot}, the most efficient method for generating the required correlated randomness is COT—a cryptographic primitive %redundent: that pre-allocates correlated randomness before executing OT, 
in which the sender's messages follow a fixed correlation \( \Delta \). Specifically, the sender holds two values \( v \) and \( v \oplus \Delta \), and the receiver holds a choice bit \( u \in \{0,1\} \) and the corresponding correlated value \( r_u \). In the input-dependent phase, the receiver masks their actual selection bit \( b \) by computing \( c = b \oplus u \) and sends \( c \) to the sender. The sender then returns the masked messages \( m_0 \oplus r_c \) and \( m_1 \oplus r_{c \oplus 1} \). Using \( r_u \), the receiver recovers \( m_b \) while learning nothing about \( m_{1 \oplus b} \). 

\subsection{Related work and Key Motivations}

The constitution of COT is crucial but not trivial. If COTs are constructed purely using public-key cryptographic techniques, each COT instance require expensive public-key operations, leading to \( O(\lambda) \) communication and computation per transfer, where \( \lambda \) is the security parameter (typically 128 or 256 bits). State-of-the-art COT generation protocols typically follow a two-stage template: (1) a setup phase that generates sparse correlations using primitives such as Homomorphic Secret Sharing (HSS), Distributed Point Functions (DPF)~\cite{boyle2019efficient1}, or OTs~\cite{yang2020ferret}; and (2) a non-interactive linear expansion phase based on the Learning Parity with Noise (LPN) assumption~\cite{boyle2019efficient, boyle2019efficient1, yang2020ferret, couteau2021silver, boyle2022correlated, roy2022softspokenot, couteau2024quietot}. While these designs are efficient on powerful machines, they are ill-suited to resource-constrained environments such as IoT sensors and wearables. %and medical devices.
% --> resource-constrained devices, e.g., xxx (real-world application devices). "embedded IoT devices" could have moderate resources

In practice, secure operations in PPMLaaS—such as ReLU and Maxpooling—rely on large volumes of COTs; however, on resource-constrained devices, these operations must be split into small batches, requiring repeated COT generation with fresh correlations for each batch. In such scenarios, although SOTA protocols claim non-interactive extension, the sparse correlation setup still incurs significant communication overhead. For example, recent silent OT-extension protocols~\cite{boyle2019efficient1} require nontrivial key exchanges—typically via DPFs—before parties can expand to a large number of OTs. Ferret~\cite{yang2020ferret} generates the required correlations using multi-point COT, which introduces considerable communication costs. Similarly, Boyle et al.~\cite{boyle2019efficient} use two rounds of OT-based key exchange, relying on Puncturable Pseudorandom Functions (PRFs) with position-specific constraints to construct sparse correlations. SoftSpokenOT~\cite{roy2022softspokenot} applies a related strategy over larger fields, where each correlation block is instantiated using a specialized OT that also demands substantial setup. Despite being labeled “silent,” these protocols still require communication in the correlation setup, becoming a performance bottleneck in resource-constrained, batch-based settings.

\section{Design of Silentflow}\label{Sec:Design}

This section presents the core design of SilentFlow, starting with a high-level architectural overview. \S\ref{sec:TEE-asssited} describes how TEEs are leveraged to eliminate interaction during COT generation. Finally, we introduce a hardware accelerator tailored to overcome the computational bottleneck in COT generation. SilentFlow assumes at most one malicious party, achieving malicious security for COT generation via TEE-synchronized randomness, while inheriting semi-honest security for online inference from the underlying MPC framework.

\begin{figure}[h!]
  \centering
  \includegraphics[width=\linewidth]{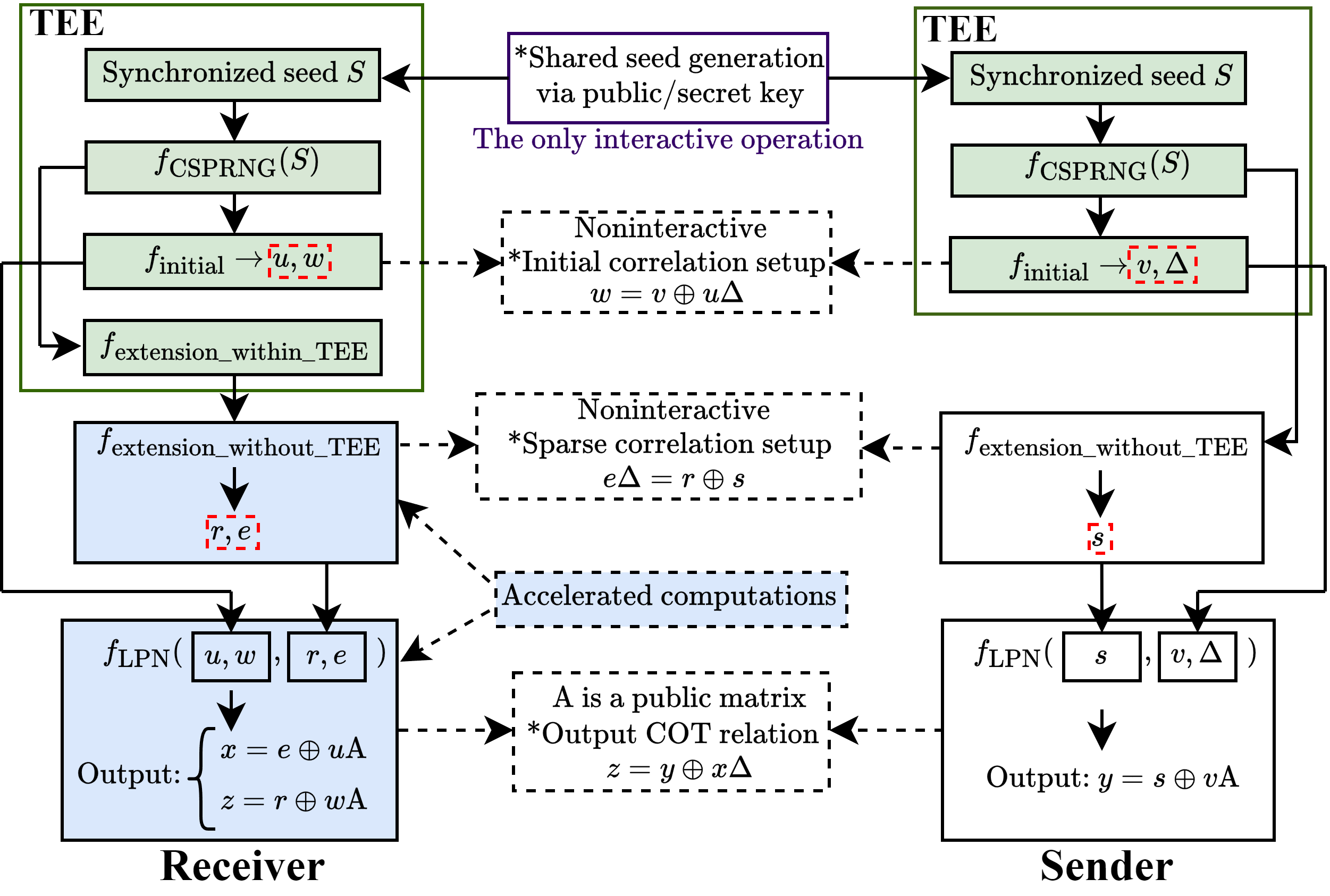}
  \vspace{-0.2in}
  \caption{System Overview of SilentFlow}
  \label{fig:system_overview}
\end{figure}

\subsection{System Architecture}

As illustrated in Fig.~\ref{fig:system_overview}, SilentFlow tackles excessive communication overhead in COT generation %addresses the above challenges 
through four integrated modules: \emph{shared seed generation}, \emph{initial correlation setup}, \emph{sparse correlation constitution}, and \emph{LPN-based local computation}. The first two modules are one-time setup steps; regardless of how many future batches are needed, execution can resume directly from the third module. Together, these components eliminate client-server communication during COT generation, supporting efficient small-batch execution without repeated network interaction.

SilentFlow initially establishes correlations of the form \(\mathbf{w} = \mathbf{v} \oplus \mathbf{u}\Delta\) non-interactively within the TEE, thereby removing costly base OTs and interaction overhead typical in state-of-the-art protocols such as Ferret~\cite{yang2020ferret}. Subsequently, the COT extension module expands these initial correlations locally in the untrusted domain, avoiding the frequent and expensive cross-boundary operations required by prior TEE-based approaches~\cite{dong2023poster}. Crucially, this non-interactive extension generates local correlations in the form of \(\mathbf{r} \oplus \mathbf{s} = \mathbf{e}\Delta\), where \(\mathbf{e}\) is a sparse binary vector (often Hamming weight). In the final step, the sparse correlation is added to the product of the initial correlation and a public matrix \( \mathbf{A} \), ensuring compliance with the LPN assumption and thereby satisfying the security requirements for COT generation. 

\subsection{TEE-Assisted COT Generation}\label{sec:TEE-asssited}

\noindent\textbf{Shared Seed Generation.} We adopt a synchronized seed generation protocol, similar to the secure initialization in~\cite{zhou2022ppmlac}, where both parties contribute random values and compute the shared seed as their sum. This ensures mutual unpredictability while allowing both parties to derive consistent CSPRNG outputs. To securely execute this protocol, secret keys and local randomness must remain confidential. Therefore, all cryptographic operations are performed within TEEs, which provide isolated execution and prevent leakage even in the presence of an untrusted client or server.

\noindent\textbf{Initial Correlation Setup.\label{sec:initial_setup}} As a synchronized seed is generated, both the sender's and receiver's TEEs deterministically sample the same vectors \( \mathbf{u} \in \mathbb{F}_2^k \), \( \mathbf{v} \in \mathbb{F}_{2^k}^k \), and the global key \( \Delta \in \mathbb{F}_{2^k} \) via CSPRNG~(Fig.~\ref{fig:system_overview}). The receiver’s TEE locally computes \( \mathbf{w} = \mathbf{v} \oplus \mathbf{u} \Delta \). The sender then sends \( \mathbf{v} \), \( \Delta \) out of TEE, while the receiver sends \( \mathbf{u} \), \( \mathbf{w} \) out of TEE. 
% The sender’s TEE then sends \( \mathbf{v} \) and \( \Delta \), while the receiver’s TEE sends \( \mathbf{u} \) and \( \mathbf{w} \), all to be used outside the TEE. 

This setup phase avoids the interactive communication required in traditional base OT protocols, reducing initialization cost and enabling fast, communication-free correlation setup. Although CSPRNGs in TEEs incur higher computational latency than lightweight PRNGs, the overhead is minimal, as only a small number of base COTs are needed to bootstrap the later extension phase. Moreover, this one-time cost is amortized over large batches of extended COTs. Compared to prior TEE-based frameworks~\cite{dong2023poster} that generate all COTs inside the enclave, SilentFlow minimizes trusted-side computation by restricting TEE usage to the initial seed and base COTs. %Additionally, fixed values like \( \Delta \) can be reused across batches, improving cache locality and reducing memory transfer overhead during extension and LPN phases.

\begin{figure}[h!]
  \centering
  \includegraphics[width=\linewidth]{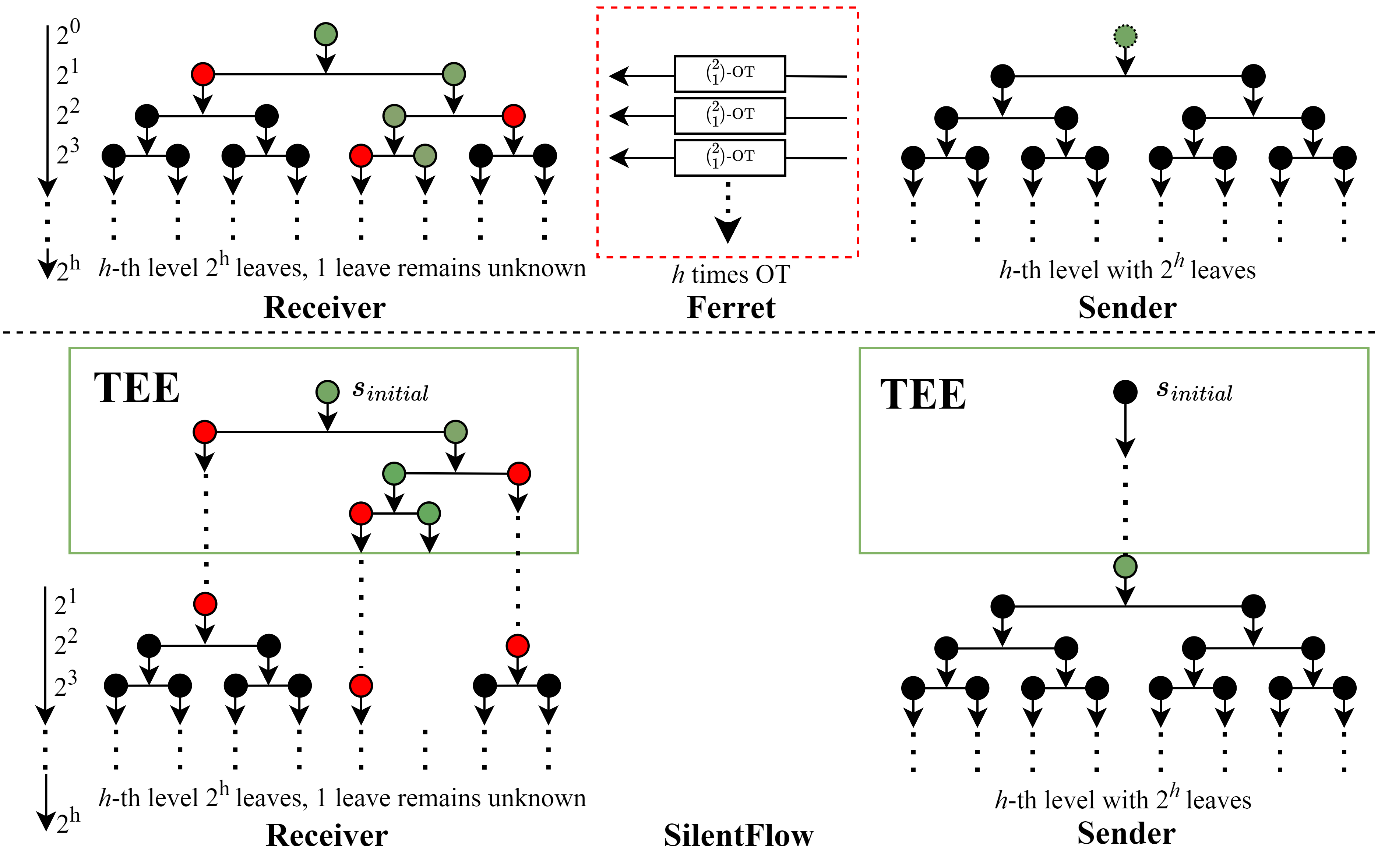}
  \captionsetup{skip=-15pt}
  \caption{Design of non-interactive sparse correlation constitution (GGM extension)}
  \label{fig:noninteractive_extension} 
\end{figure}

\noindent\textbf{Non-interactive Sparse Correlation Constitution.\label{sec:extension}} As shown in Fig.~\ref{fig:noninteractive_extension}, we adopt the standard approach of expanding a Goldreich-Goldwasser-Micali (GGM) binary tree to construct the sparse correlation, but with a key distinction—our TEE-assisted protocol enables this process to be entirely non-interactive, a feature not achieved by state-of-the-art works~\cite{boyle2018compressing,boyle2019efficient,boyle2019efficient1,couteau2021silver}.
The constitution involves the receiver obtaining a set of seeds from the sender, where the seed corresponding to the secret index is hidden, and the remaining \textit{red} seeds are extended to the leaves of the tree. To ensure that the receiver obtains only one of two seeds at each tree level, protocols like Ferret~\cite{yang2020ferret} perform a 1-out-of-2 OT at every level of the GGM tree. As the receiver traverses this single secret path using the OT-derived seeds, it reaches the \( h \)-th level and reconstructs all leaves except the one corresponding to the nonzero entry in \( \mathbf{e} \). This process guarantees that the sender and receiver generate correlated pseudorandom vectors \( \mathbf{s} \) and \( \mathbf{r} \) satisfying \( \mathbf{r} + \mathbf{s} = \mathbf{e} \cdot \Delta \). However, this method incurs high communication overhead under resource-limited devices due to low-batch execution, requiring \( h \) OT operations per extension, which becomes a major performance bottleneck. %in resource-constrained environments.

Our key idea is to replace the costly OT-based selection with a synchronized, shared seed within the TEE. Specifically, $S_{\text{TEE}}$ and $R_{\text{TEE}}$ generate the same initial seed $s_{\text{initial}}$ on each side by following the steps outlined in the shared seed generation. $S_{\text{TEE}}$ send $s_{\text{initial}}$ directly to the untrusted domain for tree extension, while $R_{\text{TEE}}$ retains $s_{\text{initial}}$ to keep it hidden from the untrusted environment. On the sender's side, the extension is systematic, involving the expansion of $s_{\text{initial}}$ to perform $2^{h+1}$ computations up to level $h$. 

The main challenge lies in extending only the selected seed without revealing its identity to the receiver. To address this, $R_{\text{TEE}}$ generates a single value \( b \in \{0, 1\}^h \) that encodes the selection bits across the tree levels, where the path defined by \( b \) corresponds to either the red or green seed. For each extension at level \( i \in \{1, \ldots, h\} \), $R_{\text{TEE}}$ retains the green seed along the path determined by \( b \), while the complementary red seed is released to the untrusted domain to enable faster seed expansion. As a key optimization of the TEE-assisted extension compared to prior work~\cite{dong2023poster}, only \( h \) seed expansion operations are performed inside the TEE, while the remaining \( 2^{(h+1)} - h \) seed expansions are offloaded to the untrusted domain, inducing only \( h - 1 \) trust boundary crossings. This avoids costly computation inside the TEE. At each level, the green seed on the secret path remains hidden from the receiver. At the final level, a masked value is computed inside $R_{\text{TEE}}$ by adding the global key to the corresponding green leaf. This value is then released to the receiver. As a result, the receiver learns only the masked value and the unmasked red values, while the secret leaf remains hidden. This achieves the same effect as the communication-intensive OT-based approach used in Ferret~\cite{yang2020ferret}, but with only \( h - 1 \) trust boundary crossings and no interaction between the client and server. For clarity, the figure illustrates a single GGM tree extension, though multiple trees are processed in parallel in practice.

\subsection{Hardware Acceleration of Silentflow}\label{sec:LPN-GGM}

This section introduces a novel hardware–algorithm co-design aimed at accelerating COT generation (particularly, sparse correlation constitution and LPN-based local computation) in Silentflow. Using the Roofline model~\cite{roofline}, which relates computational throughput to memory bandwidth and arithmetic intensity, we identify that our baseline algorithm is memory-bound, indicating that performance is limited primarily by memory access rather than computation. To address this, we propose a unified optimization strategy, termed \textbf{structured algorithmic decomposition}, which targets latency bottlenecks at both the system and module levels. Our approach improves computational throughput while significantly reducing slow, energy-intensive off-chip data transfers~\cite{mmm}.

\begin{figure}[!h]
  \centering
  \includegraphics[width=\linewidth]{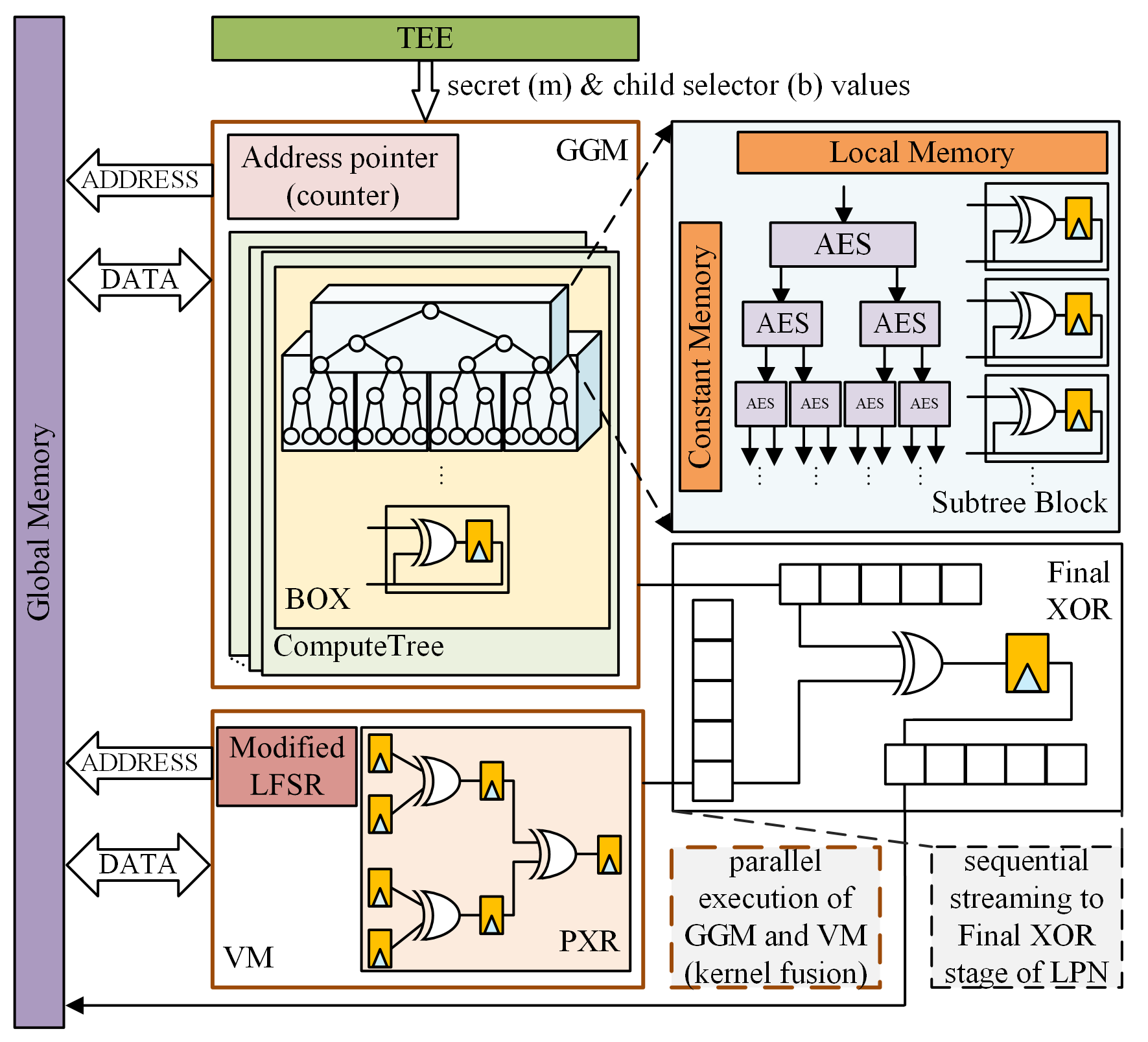}
  \captionsetup{skip=-15pt}
  \caption{System Decomposition of SilentFlow}
  \label{fig:system_diagram2} 
\end{figure}

\noindent\textbf{Latency Optimization via Kernel Fusion.}\label{sec:SYS}The non-interactive extension phase consists of two serialized phases: GGM tree expansion (for sparse correlation) and LPN computation, typically serialized due to presumed data dependencies. However, our key insight is that the only true dependency lies in the final XOR stage. We thus apply kernel fusion to reorganize the pipeline by decoupling the LPN into two independent sub-stages: a vector-matrix multiplication (VM) and a final XOR. This decoupling transforms the original latency equation from a strictly sequential model \eqref{eq:overall_latency} to a latency-balanced parallelizable form \eqref{eq:par_latency}.
\vspace{-0.05em}
\begin{equation}
    \text{Latency} = L_{\text{GGM}} + L_{\text{LPN}}
    \label{eq:overall_latency}
\end{equation}
\begin{equation}
    \text{Latency} = \max(L_{\text{GGM}}, L_{\text{VM}}) + L_{\text{XOR}}
    \label{eq:par_latency}
\vspace{-0.05em}
\end{equation}

To minimize latency, our design balances the computational resources between the GGM and VM modules. While GGM is inherently more compute-intensive, the runtime of both modules depends heavily on resource allocation. By employing structured algorithmic decomposition, we explore an optimal resource allocation strategy to balance latency and resource usage. This decomposition is applied not only at the system level but also recursively within each module, such as the blocked expansion used for GGM discussed in Section~\ref{sec:GGM}. Fig.~\ref{fig:system_diagram2} illustrates our hardware architecture, highlighting how structured decomposition improves bandwidth utilization, parallelism, and overall execution latency.

\noindent\textbf{GGM Acceleration via Blocked On-chip eXpansion (BOX).}\label{sec:GGM}
Each node expansion in the GGM tree uses a pseudorandom generator (PRG), typically AES in ECB mode~\cite{yang2020ferret}. To generate the required $n$ pseudorandom leaves for LPN noise inputs, the baseline method expands multiple binary trees of height $h$. This expansion is memory-intensive due to frequent global accesses to temporary intermediate nodes, resulting in poor locality, low computational intensity, and exponential memory growth with tree depth $h$ (see Table~\ref{tab:mem_cmp}).

\begin{table}[h!]
\caption{Memory access cost comparison: naive vs. proposed. $C_g$/$C_l$ denote average global/local access cycles.}
\centering
\renewcommand{\arraystretch}{1.2}
\begin{tabular}{cccc}
\hline
\textbf{Version} & \textbf{Global} & \textbf{Local} & \textbf{Cost} \\
\hline
Naive & $6 \times 2^{h-1}$ & $0$ & $6 \times 2^{h-1} \cdot C_g$ \\
\hline
Proposed & $2 \times 2^{h-1}$ & $4 \times 2^{h-1}$ & $2 \times 2^{h-1} \cdot (C_g + 2C_l)$ \\
\hline
\label{}
\end{tabular}
\label{tab:mem_cmp} 
\end{table}

To mitigate this performance bottleneck, we propose a second-level decomposition strategy, \textbf{Blocked On-chip eXpansion (BOX)}, which decomposes the GGM expansion into independent subtree blocks of depth $s$. Each subtree reads its root from global memory, expands locally to generate $2^s$ leaves, and keeps all intermediate states on-chip. This data localization enables parallel execution of subtrees at the same level, significantly improving reuse and reducing intermediate memory transfers as shown in Table~\ref{tab:mem_cmp}. Since average global memory access cost ($C_g$) significantly exceeds the local memory access cost ($C_l$), our approach reduces memory access overhead by approximately a factor of three.

\begin{figure}[!h]
  \centering
  \includegraphics[width=\linewidth]{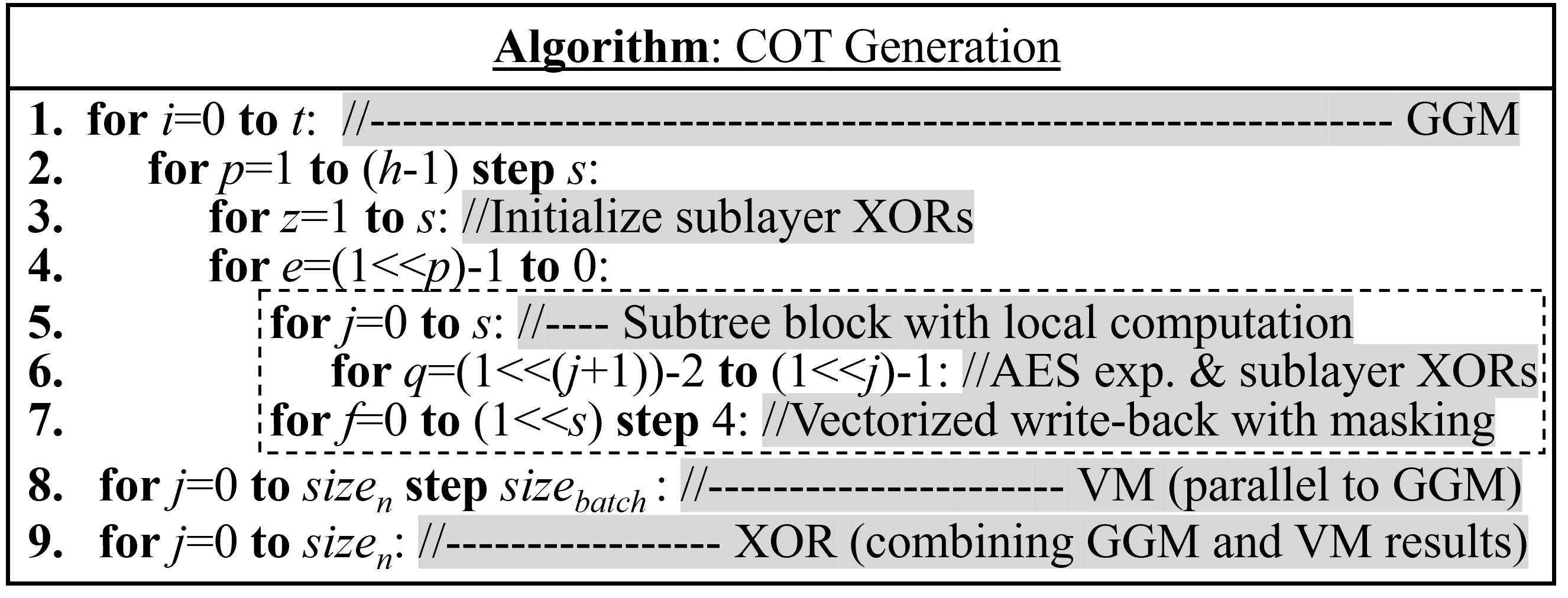}
  \captionsetup{skip=-15pt}
  \caption{Pseudocode of the proposed decomposition algorithm}
  \label{fig:list2} 
\end{figure}

As shown in Fig.~\ref{fig:list2}, our hardware unrolls inner loops to exploit fine-grained parallelism and pipelines outer loops of sub-tree block calls to achieve high-throughput dataflow. To avoid redundant global memory accesses at the final depth, masking is applied during leaf node write-back. This optimization leverages the lower latency of TEE-based data access compared to OT, which incurs higher cryptographic and network overhead. Consequently, subtrees can proceed without stalling on OT data. To identify the optimal subtree depth, we explore the design space to balance LPN and GGM latencies, aiming to minimize both absolute latency and latency gap. As shown in Fig.~\ref{fig:dse}, while $batch\_size{=}16$, $s{=}3$ achieves the smallest latency gap, $batch\_size{=}256$, $s{=}4$ offers lower absolute latency with a small gap, providing the best trade-off used in our evaluation.

\begin{figure}[t]
\captionsetup{skip=-5pt}
\centering
\begin{tikzpicture}
\begin{axis}[
    title={$2^{14}$ COT Generation},
    title style={font=\footnotesize, yshift=-1ex},
    font=\footnotesize,
    width=0.75\linewidth,
    height=0.4\linewidth,
    xlabel={Latency (ms)},
    ylabel={\textcolor{blue}{Subtree height \tikz[baseline]{\node[draw=blue, line width=0.8pt, circle, inner sep=1.3pt]{}}}},
    yticklabel style={color=blue},
    ytick style={color=blue},
    axis y line*=left,
    axis x line*=bottom,
    legend style={
        at={(0.97,0.85)},
        anchor=north east,
        legend cell align=right,
        nodes={right},
        draw=none,
        fill=none,
        row sep=5pt,
    },
    grid=major,
    xmin=0.6, xmax=1.8,
]
\addplot[
    color=blue,
    mark=o,
    thick,
]
coordinates {
    (1.707, 3)
    (1.156, 4)
    (0.682, 6)
    (0.602, 12)
};
\end{axis}
\begin{axis}[
    font=\footnotesize,
    width=0.75\linewidth,
    height=0.4\linewidth,
    ylabel={\textcolor{green!50!black}{Batch size \tikz[baseline]{\node[draw=green!50!black, fill=green!50!black, rectangle, inner sep=2pt]{}}}},
    yticklabel style={color=green!50!black},
    ytick style={color=green!50!black},
    axis y line*=right,
    axis x line=none,
    xmin=0.6, xmax=1.8,
]
\addplot[
    color=green!50!black,
    mark=square*,
    dashed,
    thick,
]
coordinates {
    (1.687, 16)
    (1.461, 32)
    (1.394, 64)
    (1.279, 128)
    (1.238, 256)
};
\end{axis}
\end{tikzpicture}
\caption{GGM and VM Latency across varying subtree heights and batch sizes}
\title{$2^{14}$ COT Generation}
\label{fig:dse} 
\end{figure}

% sub-tree modules are time-multiplexed

\begin{figure}[h]
  \centering
  \includegraphics[width=\linewidth]{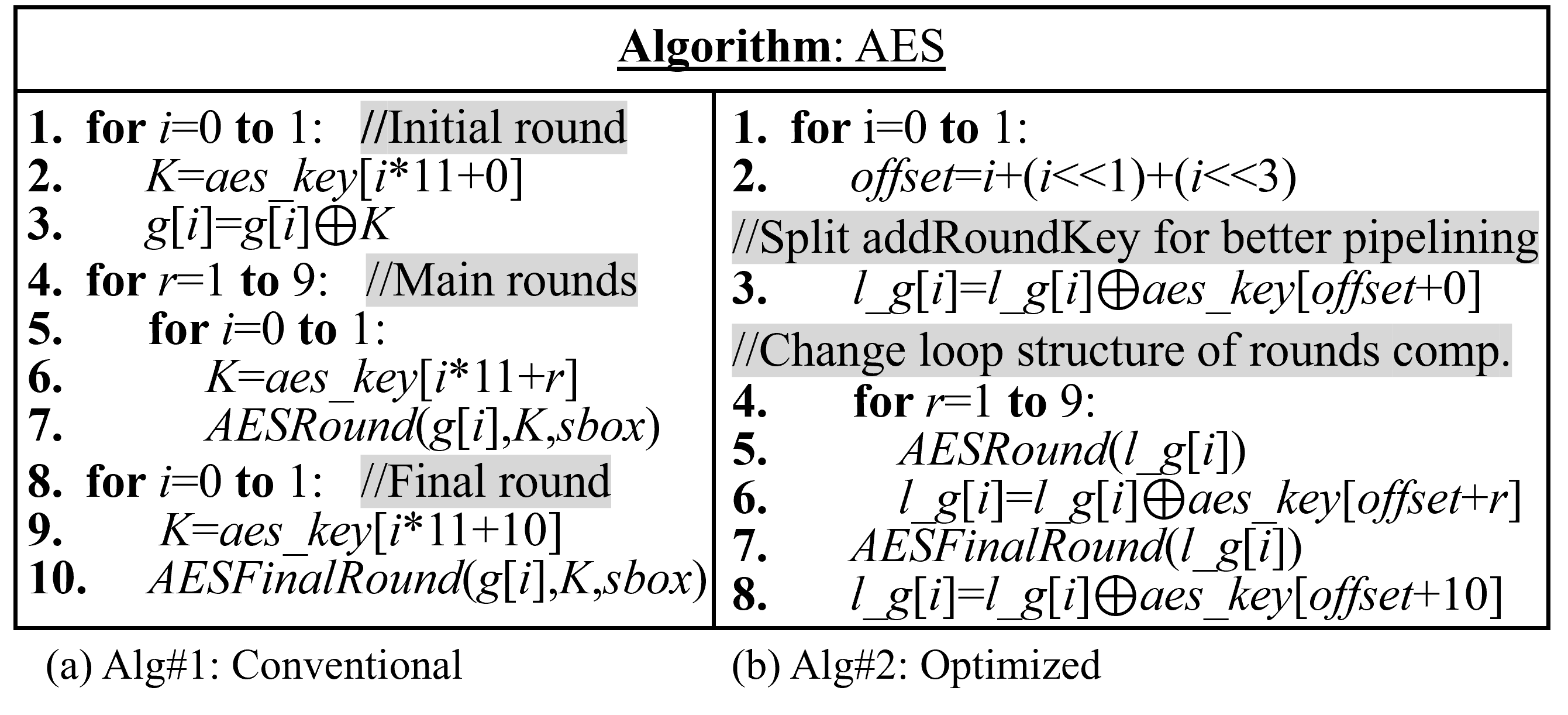}
  % \captionsetup{skip=-18pt}
  \vspace{-0.25in}
  \caption{AES expansion algorithm comparison}
  \label{fig:aes_Algorithm} 
  \vspace{0.1in}
\end{figure}

To further accelerate BOX execution, we optimize AES at the microarchitectural level, as its frequent invocation during node expansion significantly affects performance. As illustrated in Fig.~\ref{fig:aes_Algorithm}, precomputed round keys and S-box values are stored in local RAM to reduce memory overhead and eliminate redundant key expansions. The round key addition is deferred to the final stage to minimize data dependencies and enhance pipelining. Moreover, the deeply nested loop is restructured into a pipeline-friendly format, shortening critical paths and enabling efficient memory transfers across iterations.

% \vspace{-1em}

\noindent\textbf{VM Acceleration via Vectorized Batch XOR Reduction.}
The LPN computation involves a vector-matrix multiplication (VM), e.g., $\mathbf{(u,v,w)} \cdot \mathbf{A}$ in Fig.~\ref{fig:system_overview}, step~\#3, where $\mathbf{A} \in \mathbb{F}_2^{k \times n}$ is a sparse binary matrix with $d$ non-zero entries per column. Each VM computes the XOR of $d$ selected input indices, as detailed in ~\cite{boyle2018compressing,yang2020ferret}. The resulting low spatial locality—e.g., limited reuse of nearby memory addresses—leads to poor cache utilization, as only small portions of cache lines are accessed~\cite{pim}. This increases memory pressure in large-scale COT generation, where frequent global memory accesses exceed cache capacity. Prior works use AES with SIMD extensions to generate random indices, but XORs are computed sequentially, incurring two global memory $reads$ and one $write$ per operation. This read-modify-write pattern introduces loop-carried dependencies, preventing full pipelining. Linear Feedback Shift Registers (LFSRs) provide lightweight pseudorandom number generation but are also sequential, limiting parallelism in high-throughput designs.

\begin{figure}[h]
   \centering
   \includegraphics[width =\linewidth]{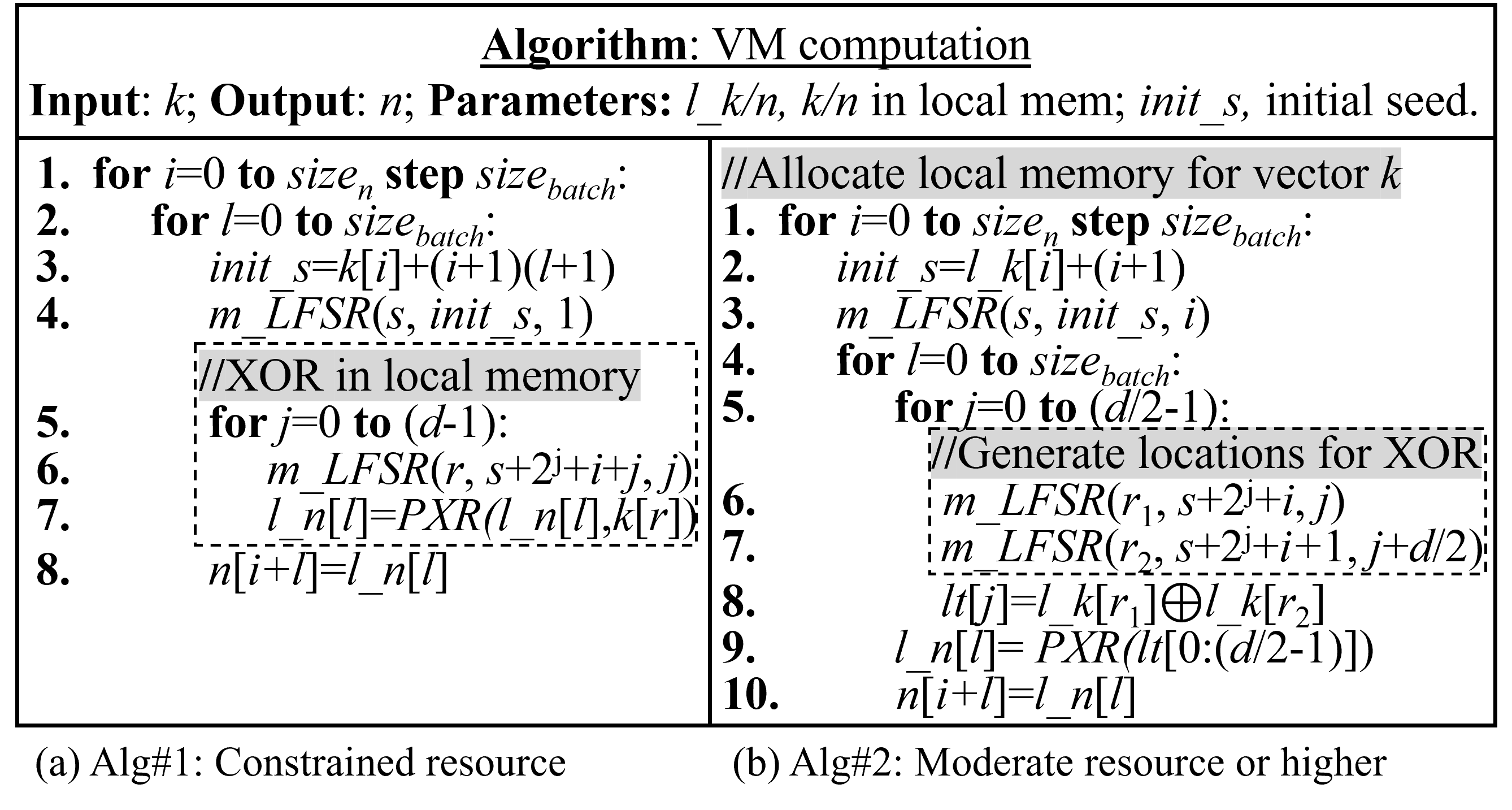}
   \captionsetup{skip=-18pt}
   \caption{Algorithms for VM computation with different resources}
   \label{fig:vm_algorithm} 
\end{figure}

To address memory constraints %and improve performance 
on resource-limited hardware, we introduce a pipeline- and parallelism-friendly approach based on fine-grained decomposition of the VM computation—decoupling index generation, memory access, and XOR reduction into independently optimized stages. Let $\mathbf{k}$ denote the input vector from the TEE-assisted Initial Correlation Setup, and $\mathbf{i} \in \mathbb{F}_k^d$ represent a set of randomly generated indices used to select entries from $\mathbf{k}$. The XOR of selected elements, $\mathbf{k}[i_1], \dots, \mathbf{k}[i_d]$, is computed in local memory, thereby reducing on-chip usage by avoiding storage of unused portions of $\mathbf{k}$%, as shown in Fig.~\ref{fig:challenges_solution} 
. We introduce two variants of a \textit{Pipelined/Parallel XOR Reducer} (PXR) that replace sequential XOR operations. The pipelined reducer (Alg\#1) avoids intermediate reads for each update, while the parallel reducer (Alg\#2) eliminates iteration dependencies, maximizing performance when additional resources are available, as shown in Fig.~\ref{fig:vm_algorithm}.

Unlike prior methods, our algorithm supports \textit{batched memory transactions} by removing LFSR feedback dependencies. We achieve this via a multiplexer that injects an auxiliary bit into the XOR path, allowing independent iterations with unique seeds derived from the iteration index. The system uses a 512-bit data bus to fetch four 128-bit elements per transaction, aligning with DDR4/AXI burst sizes for improved bandwidth and latency. Consequently, our VM module supports dataflow parallelism through pipelining, local memory reuse, and batched partitioning. Structured decomposition across GGM and VM modules further reduces memory traffic and latency, setting a new benchmark for COT acceleration. Detailed performance results are provided in Section~\ref{sec:FPGA_result}.

\section{Experiments}\label{Sec:Simu}
\subsection{Experiment Setup}

We evaluate our benchmarks under three network configurations: (1) a regular LAN environment with 3Gbps bandwidth and 0.3ms latency, (2) a constrained WAN setting with 200Mbps bandwidth and 50ms latency, and (3) a mobile-like setting with 100Mbps bandwidth and 80ms latency. For secure computation primitives, we synthesize and implement COT generation on a low-end Zynq-7000 SoC FPGA using Vitis High-Level Synthesis~\cite{vitis_hls_user_guide}. We define the client profile as a constrained configuration equivalent to typical IoT sensors (500MHz single-core CPU, 256MB mem), reflecting real-world platforms like the BeagleBone embedded system used in cryptographic benchmarking~\cite{ulcl2015}. The server resource profile reflects an edge-server setup with equivalent resources as Intel NUC 12 Pro. Intel SGX~\cite{costan2016intel} is used as the TEE testbed, Silentflow relies on a limited set of CSPRNG and PRNG operations, thereby functioning with ultra-lightweight resource requirements and avoiding computational overhead.

\subsection{FPGA Acceleration of SilentFlow}\label{sec:FPGA_result}

Table~\ref{tab:FPGA_COT} shows hardware utilization and timing results for the GGM and VM units, as well as overall COT generation, after place-and-route. The number of BRAMs reflects on-chip memory use, similar to CPU cache. We use Ferret as the baseline for comparison. Our BOX approach achieves a $6.20\times$ to $23.13\times$ speedup, depending on the subtree height configuration. For VM, our vectorized batch PXR method achieves $3.18\times$ and $5.06\times$ lower latency using Alg\#1 and \#2, respectively, with $batch\_size{=}256$—the best-performing setting in Fig.~\ref{fig:dse}. Finally, the integrated end-to-end COT generation improves latency by $10.21\times$ and $11.78\times$, with the chosen configuration balancing performance and resource trade-offs between VM and GGM, as discussed in Section~\ref{sec:LPN-GGM}. However, with sufficient BRAM, we observe up to a $53.29\times$ improvement in VM computation, highlighting BRAM’s advantage in supporting highly parallel operations. This result suggests that with adequate resources—such as high-performance embedded processors or smartphones (e.g., Raspberry Pi 4, or iPhone X and later as the client)—significant speedups are achievable. Our design thus provides two options: one for resource-limited devices and another for more capable platforms. 

\begin{table}[htbp]
  \centering
  \setlength{\tabcolsep}{2.6pt}
  \caption{Hardware resource cost of VM, GGM, and COT computation, with $k=32771,n=2^{20}$, using different optimization methods on Zynq-7000 SoC Z-7045. *Alg\#2 exceeds the FPGA storage limits and is viable when sufficient resources are available.}
  \begin{tabular}{m{2.2em}m{4.1em}cccccccc}
    \toprule
    \textbf{Unit} & \textbf{Method} & \textbf{BRAM} & \textbf{DSP} & \textbf{FF} & \textbf{LUT} & \textbf{Latency(ms)} & \textbf{Speedup} \\
    \midrule
    \multirow{5}{*}{GGM} & Ferret & 48 & 0 & 14564 & 19206 & 281.40 & - \\
                         & BOX({\scriptsize$s{=}3$}) & 67 & 0 & 31446 & 20193 & 109.37 & $6.20{\times}$ \\
                         & BOX({\scriptsize$s{=}4$}) & 67 & 0 & 42346 & 28269 & 73.99 & $9.16{\times}$ \\
                         & BOX({\scriptsize$s{=}6$}) & 67 & 0 & 82378 & 55921 & 43.82 & $15.47{\times}$ \\
                         & BOX({\scriptsize$s{=}12$}) & 227 & 0 & 73189 & 60490 & 29.31 & $23.13{\times}$ \\
    \midrule
    \multirow{4}{*}{VM} & Ferret & 4 & 0 & 8520 & 8265 & 281.40 & - \\
                        & Alg\#1 & 4 & 1 & 8748 & 3885 & 88.46 & $3.18{\times}$ \\
                        & Alg\#2 & 276 & 0 & 14272 & 13275 & 55.66 & $5.06{\times}$ \\
                        & Alg\#2* & 2072 & 0 & 25920 & 23005 & 5.28 & $53.29{\times}$ \\
    \midrule
    \multirow{3}{*}{COT} & Ferret & 84 & 0 & 22881 & 23986 & 985.79 & - \\
                         & Alg\#1({\scriptsize$s{=}4$}) & 71 & 1 & 50044 & 31870 & 96.54 & $10.21{\times}$ \\
                         & Alg\#2({\scriptsize$s{=}6$}) & 295 & 0 & 101376 & 62341 & 83.67 & $11.78{\times}$ \\
    \bottomrule
  \end{tabular}
  \label{tab:FPGA_COT}
\end{table}

\subsection{COT generation}\label{sec:cot_generation}

We compare SilentFlow with representative state-of-the-art OT protocols, as summarized in Table~\ref{tab:cot_comparison}. SilentFlow achieves substantial speedups ranging from 5.14$\times$$\sim$39.51$\times$ across different protocols. Notably, the speedup increases as network latency worsens, while SilentFlow remains unaffected due to its fully non-interactive design—even during the sparse correlation generation phase, enabled by our TEE-based approach. Consequently, the total COT generation time is dominated solely by local computation, which is further accelerated by our FPGA-based hardware architecture.

\begin{table}[htbp]
  \centering
  \caption{Performance comparison with various state-of-the-art COT generation protocols. Following the standard in prior works, the generation of $10^7$COTs is used as the benchmark, including both the sparse correlation constitution and the extension phases, while excluding the negligible one-time setup.}
  \resizebox{\columnwidth}{!}{
    \begin{tabular}{m{5em}cccccc} % Added one more column
    
    \toprule
    
     \multirow{2}[2]{*}{\parbox{5em}{\centering \textbf{Protocol}}} & \multicolumn{6}{c}{\textbf{$10^7$ COTs generation (second)}}\\ 
    
    \cmidrule{2-7}
    
    \multicolumn{1}{c}{} & \multicolumn{1}{c}{\centering LAN} & \multicolumn{1}{c}{\centering Speedup} & \multicolumn{1}{c}{\centering WAN} & \multicolumn{1}{c}{\centering Speedup} & \multicolumn{1}{c}{\centering Mobile} & \multicolumn{1}{c}{\centering Speedup} \\%&\multicolumn{1}{c}{\centering Comm.(MB)}&\multicolumn{1}{c}{\centering Round}\\
    
    \cmidrule{1-7} 

    \centering QuietOT~\cite{couteau2024quietot} & 36.24  & 29.46$\times$& 43.89 & 35.65$\times$& 48.48 &39.51$\times$\\ 
    \centering SilentOT~\cite{boyle2019efficient1} & 17.32 & 14.08$\times$ & 24.97 & 20.28$\times$& 29.56& 24.09$\times$\\ 
    \centering Ferret~\cite{yang2020ferret} & 9.703 & 7.89$\times$ & 17.02  & 13.83$\times$& 21.64&17.64$\times$\\ 
    \centering SSOT~\cite{roy2022softspokenot} & 6.33 & 5.14$\times$ & 13.98 & 11.36$\times$ & 18.57 &15.13$\times$\\ 
    \centering  \textbf{SilentFlow} & 1.230   &  - & 1.231 & -&1.227&-\\ 
    
    \bottomrule
    
    \end{tabular}%
  }
  \label{tab:cot_comparison}%
  %\vspace{1mm}
  %\caption*{\textit{Note:} SSOT refers to SoftSpokenOT~\cite{roy2022softspokenot}.}
  %\footnotesize \noindent\text{Note:} SSOT refers to SoftSpokenOT
  %\vspace{-0.05in}
\end{table}%

\subsection{End-to-end Framework}

In Table~\ref{tab:framework_comparison}, we compare SilentFlow against the Ferret and IKNP protocols integrated into state-of-the-art PPMLaaS frameworks. We evaluate two deep learning models: ResNet-50, representing a large-scale network, and SqueezeNet, representing a lightweight architecture. The results show that with SilentFlow, the Cheetah and CrypTFlow2 frameworks can complete inference under a mobile network environment in 60s for ResNet-50 and 152s for SqueezeNet, achieving a speedup of 4.75$\times$ to 4.78$\times$ over previous protocols. As network conditions improve, the relative speedup from communication reduction diminishes; however, SilentFlow still achieves at least a 3.17$\times$ acceleration on ResNet-50, primarily due to FPGA-based computation, as communication is typically not a bottleneck under LAN settings. These results demonstrate that although SilentFlow primarily targets COT generation, it delivers a $2.88\times$$\sim$$4.78\times$ improvement in end-to-end inference performance—validating the effectiveness of our design. 

\begin{table}[htbp]
  \centering
  \setlength{\tabcolsep}{4pt} % Adjust column separation
  \caption{Performance comparison with Ferret and IKNP protocols by implementations in real-world PPMLaaS frameworks.}
  \resizebox{\columnwidth}{!}{
    \captionsetup{skip=10pt}
    %\begin{tabular}{m{4em}m{8em}cccc} % Added one more column
    \begin{tabular}{m{8em}cccccc} % Added one more column
    
    \toprule
    
    \multirow{3}[-2]{*}{\parbox{8em}{\centering \textbf{Framework}}} & \multicolumn{3}{c}{\textbf{ResNet50 (second)}} & \multicolumn{3}{c}{\textbf{SqueezeNet (second)}}\\
    
    \cmidrule{2-7}
    
    \multicolumn{1}{c}{} & \multicolumn{1}{c}{\centering \shortstack{LAN}} & \multicolumn{1}{c}{\centering \shortstack{WAN}} & \multicolumn{1}{c}{\centering \shortstack{Mobile}} & \multicolumn{1}{c}{\centering \shortstack{LAN}}& \multicolumn{1}{c}{\centering \shortstack{WAN}}& \multicolumn{1}{c}{\centering \shortstack{Mobile}}\\
    
    \cmidrule{1-7} 
    
    \centering SCI~\cite{rathee2020cryptflow2} &  581  &  1058  &1424& 357&584&728\\ 
    \centering SCI+\textbf{SilentFlow}&183  &266  & 308 &124&139&152\\ 
    \cmidrule{1-7}
    \centering Speedup& $3.17\times$ &  $3.97\times$     & $4.62\times$   &2.88$\times$&4.20$\times$&4.78$\times$\\ 
    \cmidrule{1-7} 
    \centering Cheetah~\cite{huang2022cheetah} & 500     & 795  &1013& 129 &209 & 285\\ 
    
    \centering Cheetah+\textbf{SilentFlow} & 145 & 216  &256& 38 & 52 & 60\\ 
    \cmidrule{1-7}
    \centering Speedup& $3.448\times$ &  $3.68\times$    & $3.95\times$   &3.39$\times$&4.02$\times$&4.75$\times$\\ 
    
    \bottomrule
    
    \end{tabular}%
  }
  \label{tab:framework_comparison}%
\end{table}%

\section{Conclusion}\label{Sec:Summary}

In this work, we demonstrate that secure MPC-based deep learning inference is becoming practical in resource-constrained environments by addressing the primary bottleneck of COT generation. SilentFlow introduces a hardware–algorithm co-design that enables SqueezeNet inference in under 1 minute over mobile networks on hardware equivalent to IoT sensors or wearables, achieving a 4.78$\times$ speedup. For more complex models like ResNet-50, SilentFlow completes inference in under 5 minutes with a 4.62$\times$ speedup. 

{\footnotesize % Start small font size
\bibliographystyle{IEEEtran}
\bibliography{cite}
} % End small font size

\end{document}